\documentclass[prl,amssymb,amsmath,superscriptaddress,twocolumn,longbibliography,nofootinbib]{revtex4-1}

\usepackage{color,braket,tikz,calligra}
\usepackage[T1]{fontenc}
\usetikzlibrary{arrows}

\newcommand{\be}{\begin{equation}}
\newcommand{\ee}{\end{equation}}

\begin{document}

\title{Operational nonlocality}
\author{R.
  Srikanth}   \email{srik@poornaprajna.org}  \affiliation{Poornaprajna
  Institute of Scientific Research, Bengaluru 560 080, India}

\begin{abstract}
An operational concept of locality  whose quantum violation is indicated  independently of any other assumption(s) seems to be lacking  in the quantum foundations  literature so far.  Bell's theorem only shows that quantum correlations violate the conjunction of the ontological assumptions of localism  and determinism.  Taking a cue from computational complexity theory, here we define such a concept of locality in terms of a class of operational decision problems, and propose that Einstein-Podolsky-Rosen (EPR) steering  in its basic asymmetric formulation gives a specific realization of this class. Various nonclassical (convex) operational theories, including quantum mechanics, are shown to be nonlocal according to this operational criterion. We  discuss several ramifications arising from this result. It indicates a basic difference between quantum no-signaling and relativistic no-signaling, and suggests an information theoretic derivation of the former from other basic principles in the convex framework. It elucidates the connection between two different bounds on nonlocality, due to Oppenheim and Wehner (2010) and Banik et al. (2015), thereby highlighting the interplay and contrast between uncertainty and measurement incompatibility.  Our result provides an  argument supporting the realist interpretation of the quantum state, and helps clarify why  Spekkens' toy model (2007) features steering but not Bell nonlocality. 
\end{abstract}


\maketitle

\paragraph{Introduction.}
Is there an operational criterion  by which a measurement on particle $A$ can be said to disturb a distant quantum-correlated particle $B$, independently of other assumptions? While quantum information processing tasks such as remote state preparation \cite{pati1999minimum, bennett2005remote}, quantum teleportation \cite{bennett1993teleporting} and remote steering \cite{cavalcanti2016quantum} are suggestive of such a disturbance (one that, in the last case, Schr\"odinger found disconcerting \cite{schrodinger1935mathematical}), yet by virtue of intrinsic randomness in measurement outcomes, which enforces no-signaling, the  reduced   state  of $B$ is unaffected  by local actions at $A$.   This  obfuscates  the   issue  of  whether  the  remote disturbance of the state of $B$ is indeed an operational phenomenon (i.e., one that refers to operational quantities only), or merely an influence felt in a HV model reproducing the experimental observations, or yet again, just an epistemic update conditioned on Alice's measurement.  It has been the subject of some debate (cf. \cite{cavalcanti2012bell}) since the historic paper by Einstein, Podolsky and Rosen \cite{einstein1935can}.

In this context, Bell's  theorem  \cite{bell1966problem} only shows that  quantum correlations violate the conjunction of the ontological assumptions of localism and determinism, and implies nothing about localism by itself. The theorem can be formulated as the  complementarity between randomness ($I$) and signaling ($S$) resources required for simulating a violation of the CHSH inequality \cite{clauser1969proposed}, given by: 
\begin{equation}
S + 2I \ge C,
\label{eq:cmp}
\end{equation}
where $C$ is the average communication cost in bits \cite{aravinda2015complementarity, aravinda2016extending} (generalizing earlier results reported in \cite{hall2010complementary,   kar2011complementary}).  

In a deterministic HV model ($I=0$), or even certain predictively superior ($0 < I < \frac{C}{2}$) HV models, nonlocal correlations ($C>0$) imply a disturbance at the HV level ($S>0$). The operational implication of Bell's theorem is this: since special  relativity and the structure of QM impose signal locality ($S=0$), given nonlocal correlations ($C > 0$), from Eq. (\ref{eq:cmp}), we infer non-vanishing operational randomness, i.e., unpredictability ($I>0$)--  a conclusion reached differently in \cite{cavalcanti2012bell}. 

As far as we know, there is thus far in the literature, no operational concept of locality that captures the notion of disturbance referenced in the question posed at the start of this Section. In this article, we propose an affirmative response to the question. 
Taking  a  leaf  from  computational complexity theory's book, we identify this concept with a class of operationally decidable problems, and propose that quantum steering realizes this class. Conceptual ramifications of our result include making a distinction between quantum and relativistic no-signaling; and making a case for no-signaling being a derived-- rather than basic-- principle in the framework of generalized  probability theories (GPTs) \cite{janotta2014generalized, hardy2011reformulating}, or convex operational theories  (Appendix I).  Quantum  mechanics (QM), formulated   as   an operational theory,  is a  special case  in this framework.  We also show that our  result  indicates  an rich interplay between  uncertainty \cite{WW10}, measurement incompatibility \cite{busch1985indeterminacy}, steering, nonlocality \cite{brunner2014bell} and no-signaling.  Our starting point takes  a  cue  from  computer  science.

\paragraph{Detectable vs. verifiable disturbance.}
Two  major  computational  complexity  classes  are  {\bf  P}, the set of all decision problems  that are  quickly  (i.e., in  time that  scales polynomially with problem input size) solvable, and {\bf  NP}, the set of problems that  are quickly verifiable (against a certificate).  Obviously,  ${\bf P} \subseteq {\bf NP}$, but whether the containment is strict is the {\bf P} versus {\bf  NP} problem,  a major  open problem  in computer  science \cite{aaronson2013quantum}. 

Here we consider an analogous use of the concept of verification against a certificate to capture the idea that although the putative remote disturbance is not detectable (thanks to no-signaling), it may be checkable \textit{a posteriori} in an operational way. In the context of bipartite correlations,  Bob's measurement is considered to be the analogue of  polynomial-time computation, and his guessing of Alice's measurement choice with better-than-random probability based on his outcome, as the analogue of problem solving.  The analogues of decision problems are operational theories, and those of problem instances are individual (bi-partite) states.

Accordingly,   \textbf{Det}-- the class of signaling theories, i.e., ones wherein Bob can (with better than random probability) unilaterally determine Alice's input--  is analogous to \textbf{P} (or \textbf{BPP}, but this distinction isn't significant here).   The analog of \textbf{NP} would be  the class of  theories wherein Bob can operationally verify (through some procedure $\mathcal{P}$) Alice's action against a certificate subsequently issued by her, which (class) we designate \textbf{Ver}. Clearly, $\textbf{Det} \subseteq \textbf{Ver}$. 

If (over many trials) a state $\omega_{AB}$ can pass the verification test $\mathcal{P}$  for some measurement setting(s),  then $omega_{AB}$ is deemed \textit{operationally nonlocal}. A theory that features any such state belongs to \textbf{Ver}. Since remote disturbance is trivial to operationally verify in signaling theories,  our first concern is to identify $\mathcal{P}$ suitable for ``\textbf{Ver}-complete theories''-- i.e., non-signaling theories in \textbf{Ver}.  In what follows, GPTs are implicitly assumed to be non-signaling, unless explicitly stated to be otherwise. 

 Theories in $\overline{\bf Ver}$, the complement of \textbf{Ver} in the space of all GPTs, are, by definition, operationally local (or, ``Einstein local''). They correspond to the classical (and strongest) notion of locality.  By contrast, theories in $\overline{\bf Det}-\overline{\textbf{Ver}}$ are local in a weaker sense. They violate operational locality and correspond to the quantum or nonclassical notion of locality.
 

In general, theories in \textbf{Det} will lack a tensor product structure, and be higher-dimensional than their non-signaling counterparts in \textbf{Ver}. For example, in the context of theories of finite-input-finite-output bipartite  correlations  $P(ab|xy)$, where  $x  \in \mathcal X$ and $a  \in \mathcal A$ (resp., $y \in  \mathcal Y$ and $b \in \mathcal  B$) are the  inputs and  outputs of Alice  (resp., Bob), a theory in \textbf{Det} is larger dimensional than its non-signaling counterpart by    (Appendix II):
$$
C_{\rm nosig} = |\mathcal{X}| (|\mathcal{Y}|-1) (|\mathcal{A}|-1) +  |\mathcal{Y}| (|\mathcal{X}|-1)(|\mathcal{B}|-1),
$$
the number of independent  no-signaling  constraints.

Being signaling and hence incompatible with special relativity, theories in \textbf{Det} will be considered unphysical. However, as would be clear at the end of this Article, it is advantageous to widen the GPT framework to include them, in order to better understand what principles single out QM as a special non-signaling theory in Nature. 

\paragraph{Operational nonlocality.}

Alice and  Bob live in a world whose  physical laws are governed by GPT $ \theta $.   They have access to  well characterized state preparations and  measurements, and cooperatively implement the following steering-inspired realization of the verification test $\mathcal{P}$. Other possible realizations are discussed later.

They agree on two dichotomic incompatible measurements of Bob, $y=y_0$ and $y=y_1$, with a single-system uncertainty relation. Let  $q(y,\omega)$  denote  the probability  $\max_b  P(b|y, \omega)$ for  an arbitrary (mixed)  state $\omega$ with Bob.    A  nontrivial uncertainty  exists if, for any given state $\omega$
\begin{equation}
q(y_0) + q(y_1) \le \upsilon_{\rm loc}^\ast.
\label{eq:unc}
\end{equation}
and  $\upsilon^\ast_{\rm loc} < 2$. As  an example, in QM,  let $y_0  \equiv \sigma_X$  and $y_1  \equiv \sigma_Z$.   Then,  $\upsilon^\ast_{\rm loc}  =  1 + \frac{1}{\sqrt{2}} \approx 1.7$, with the optimal  states attaining this bound being eigenstates of  $\frac{1}{\sqrt{2}} (\sigma_X  \pm  \sigma_Z)$. Suppose Alice prepares and sends one of these states.
If Bob announces one of $y_0$ and $y_1$ randomly, then she can predict Bob's outcome with probability $\frac{\upsilon_{\rm loc}^\ast}{2}$.

Instead, suppose  Alice prepares a joint  state $\sum_{\lambda}\mathfrak{p} (\lambda)\omega_{AB}^\lambda$, with $\sum_\lambda \mathfrak{p} (\lambda)=1$, and sends particle $B$ to Bob. After Bob announces $y_j$, Alice  measures a corresponding $x_j$  on  particle  $A$.  To prove that her action remotely disturbed Bob's state, Alice sends Bob a classical certificate predicting Bob's outcome $b_j$. Then, Bob measures $y$ and checks her claim. Over many runs, he determines the uncertainty conditioned on her certificate.

The assemblage of unnormalized states of Bob $\{\tilde{\omega}^{a|x}_B\}$ (the tilde indicating non-normalization) produced if Alice measures $x$  obtaining outcome $a$
has the  general decomposition
\begin{equation}
\tilde{\omega}^{a|x}_B = \sum_\lambda \mathfrak{p}(\lambda)
p(a|x,\lambda)\omega_B^{a|x,\lambda},
\label{eq:notlocal}
\end{equation}
where $p(a|x, \lambda)  = [e^{a|x} \otimes u_B](\omega_{AB}^\lambda)$,  $e^{a|x}$ is Alice's  effect, and $u_B$ is the identity operation on Bob's particle. 
If $\omega_{AB}^\lambda$ has the product form $\omega_A^\lambda\otimes\omega_B^\lambda$, Eq. (\ref{eq:notlocal}) reduces to:
\begin{equation}
\tilde{\omega}^{a|x}_B = \sum_\lambda \mathfrak{p}(\lambda)
p(a|x,\lambda)\omega_B^{\lambda},
\label{eq:local}
\end{equation}
where  states $\omega_B^\lambda$  are  classically correlated with $A$, with probability  $p(a|x, \lambda)$ reducing to $e^{a|x}(\omega_{A}^\lambda)$. In contrast to Eq. (\ref{eq:notlocal}), the form  Eq. (\ref{eq:local})  defines an \textit{unsteerable} assemblage  of local hidden states at $B$ \cite{wiseman2007steering}   in  the  context  of  GPTs \cite{banik2015measurement}. 
In this case, conditioning on outcome $(a|x)$ doesn't reduce Alice's uncertainty about Bob's outcomes. Therefore,  the conditioned uncertainty relation
\begin{equation}
  q(y_0|x_0) + q(y_1|x_1)
  ~\le~ \upsilon_{\rm loc}^\ast,
 \label{eq:eta}
\end{equation}
which reduces to the uncertainty relation (\ref{eq:unc}), must hold. Here, $q(y_j|x_j) \equiv \sum_{\lambda, a_j} \mathfrak{p}(\lambda)p(a_j|x_j,\lambda)q(y_j, \omega_{B}^{a_j|x_j, \lambda})$. Thus, a violation  of inequality Eq. (\ref{eq:eta}) can arise only from a steerable (from Alice to Bob) assemblage of form Eq.  (\ref{eq:notlocal}).  By convexity, to produce a violation, $\lambda$ may be fixed to be some optimal state.

 The following purely operational facts demonstrate a kind of nonlocality without invoking assumptions about a HV model. Let  $t_{\rm prep}$ be the time when Alice prepares in some unspecified way the state of particle $B$, and $t_{\rm rec}$ the time when she receives Bob's message on his chosen $y$.  (a) 
If Alice prepared the state of $B$ before knowing $y$, then $B$  is constrained by the bound of Eq. (\ref{eq:eta}), ie.,
\begin{equation}
t_{\rm prep} \le t_{\rm rec} ~\Longrightarrow~\sum_j q(y_j|x_j) \le \upsilon^\ast_{\rm   loc}.
\end{equation}
(b) The violation of Eq. (\ref{eq:eta})  would imply that  Alice prepared  Bob's state \textit{after} knowing  $y$:
\begin{equation}
\sum_j q(y_j|x_j) > \upsilon^\ast_{\rm   loc} ~\Longrightarrow~ t_{\rm prep} > t_{\rm rec} ,
\end{equation}
with no assumptions made concerning Alice's preparation method.  
(c) If Alice prepared Bob's state by measuring $A$ on (several copies of a given) state $\omega_{AB}$, and particles $A$ and $B$ are sufficiently far from each other, and further Eq. (\ref{eq:eta}) is found violated,  then it  follows that she prepared his state from afar-- implying the operational nonlocality of $\omega_{AB}$.

In other words, violation of Eq. (\ref{eq:eta}) would represent evidence of Alice's remote disturbance  of $B$ purely at   the   operational   level. By  contrast,  Bell  nonlocality  only entails an  \textit{ontic} disturbance in a class of HV models.   Operational locality of $\omega_{AB}$  is   identified  with the satisfaction of  inequality (\ref{eq:eta}) for all possible settings. Any other uncertainty-based steering inequality than Eq. (\ref{eq:eta}) can equally well be used for this argument.  

As a quantum realization of operational nonlocality, let $\omega_{AB}$ be the quantum singlet state $\frac{1}{\sqrt{2}}(\ket{01}-\ket{10})$, and Bob announces one of $y_0 = \sigma_X$ and $y_1=\sigma_Z$. Alice  measures in the   announced basis $y$,  and is able  to  predict  his outcome  with complete  certainty, leading to a violation of Eq. (\ref{eq:eta}), with the lhs being 2, whilst the rhs  $\upsilon_{\rm   loc}^\ast  \approx  1.7$. 

To violate Eq.   (\ref{eq:eta}), Alice's observables  $x=x_0$ and $x=x_1$  must be mutually incompatible. Suppose, to the contrary, they are jointly measurable. Let the corresponding outcomes be $a_0$ and $a_1$. By definition of joint measurability \cite{busch1984various, busch2013comparing}, the outcome function $p(a|x,  \lambda)$ should be  derivable as the marginal statistics of a joint probability distribution (JD) over $x_0$ and $x_1$ 
In this case,  Eq. (\ref{eq:notlocal}) becomes:
\begin{equation}
\tilde{\omega}^{a_i|x_i}_B = \sum_{\lambda} \mathfrak{p}(\lambda)
\sum_{a_{i\oplus1}}
p(a_0, a_1|x_0, x_1, \lambda)\omega_B^{a_i|x_i,\lambda}.
\label{eq:cnotlocal}
\end{equation}
where $\oplus$ signifies addition modulo 2. No-signaling implies that $\sum_{a_0} \tilde{\omega}^{a_0|x_0}_B = \sum_{a_1} \tilde{\omega}^{a_1|x_1}_B$. It follows from Eq. (\ref{eq:cnotlocal}) that 
$\sum_{a_0,a_1,\lambda}p(a_0, a_1|x_0, x_1,\lambda) (\omega_B^{a_0|x_0,\lambda}-\omega_B^{a_1|x_1,\lambda})=0$ for all states $\omega_{AB}$. This can hold true in  general only  if  for either $a_i$,  $\omega_B^{a_i|x_i,\lambda} = \omega_B^{\lambda}$, i.e., the assemblage  Eq.  (\ref{eq:cnotlocal}) takes the unsteerable form Eq. (\ref{eq:local}). In the inverse, if the $a_j$'s are incompatible, then clearly in general the unsteerable form Eq.  (\ref{eq:cnotlocal}) holds (cf. \cite{banik2015measurement, quintino2014joint,   uola2014joint, karthik2015joint}).  It is straightforward to extend this argument to more than two pairs of observables.

A subtlety here is that the JD $p(a_0,a_1 x_0, x_1, \lambda)$ may exist even though the joint measurement isn't part of the operational theory. An example is Spekkens' toy theory  \cite{spekkens2007evidence}. This peculiarity due to the non-convexity of the theory (see Appendix III). In this case, we deem the measurements to be incompatible, but ``meta-compatible'' (Appendix IV). Such a theory supports EPR steering, despite being local.

In a non-signaling theory,  Alice's ability to predict Bob's outcome is unaffected even if Bob  measures before announcing $y$.  Then, by an argument similar to that used above to show the incompatibility of the $x_j$'s, one can show that $y_0$ and $y_1$ must be mutually incompatible to violate Eq. (\ref{eq:eta}) (cf. \cite{berta2010uncertainty}).


The above observations suggest that operational nonlocality is mathematically equivalent to steerability, although the corresponding protocols have different objectives. (In the EPR steering scenario, performing unknown-to-Bob measurements, Alice aims to convince Bob, who has full control over his quantum measurements, that her state is entangled with his.) Indeed, Eq. (\ref{eq:eta}) can be considered as a steering inequality, analogous to entropic or fine-grained steering inqualities \cite{walborn2011revealing, schneeloch2013einstein, pramanik2014fine, chowdhury2014einstein}, but based on a different quantification of uncertainty and applicable to any operational theory, not just QM.   The above mentioned equivalence indeed holds good in QM, and therefore, within QM, operational nonlocality is strictly weaker than Bell nonlocality.  However, the equivalence fails in in the space of all non-signaling GPTs. In particular, PR box world \cite{barrett2005popescu, barrett2007information}, which allows perfect steering, as
evidenced by its maximal Bell nonlocality \cite{oppenheim2010uncertainty}, 
lacks operational nonlocality because pure gbits lack uncertainty, i.e., $\upsilon_{\rm loc}^\ast=1$ (Appendix III). 

Class  {\bf Ver} can be realized as the set of theories that are operationally nonlocal in the sense of allowing the violation of Eq. (\ref{eq:eta}) for some state(s). Evidently, QM  belongs to \textbf{Ver}-complete and so does Spekkens' (local) toy theory. But PR box world, for the reason stated above, doesn't, though this may not be the case in other realization of \textbf{Ver.} See  Appendix III for further discussion on the status of operational nonlocality in various theories.

\paragraph{Uncertainty, incompatibility, steering and nonlocality.}
Our realization of  operational nonlocality based on uncertainty fails to distinguish distinct theories having the same level of uncertainty bound $\upsilon_{\rm   loc}$, despite their distinct steering capabilities, e.g., classical theory and PR box world. By contrast, Bell nonlocality obviously distinguishes these two theories.


Indeed, using a bound on the violation of the CHSH inequality from measurement incompatibility \cite{banik0degree}, and a relationship between uncertainty and incompatibility for a family of GPTs  (Appendices V, VI), we obtain the Bell-CHSH inequality
\begin{equation}
|\langle  a_0b_0\rangle  +  \langle a_0b_1\rangle  + \langle a_1b_0\rangle - \langle  a_1b_1\rangle| \le 4\varsigma(\upsilon_{\rm loc}^\ast-1),
\label{eq:bellnonloc}
\end{equation}
featuring a different bound in terms of uncertainty and steering strength $\varsigma$, than that obtained in  \cite{oppenheim2010uncertainty} in terms of fine-grained uncertainty and steering. This contrasts with the case of incompatibility, which can bound Bell nonlocality by itself \cite{son2005joint, kar2016role}.  The reason is that whereas incompatibility directly relates to the (in)existence of a joint probability distribution over all inputs, uncertainty requires to be supplemented by steering strength (Appendix VI), as suggested by our above discussion.

In Eq. (\ref{eq:bellnonloc}), $\frac{1}{2} \le \varsigma \le 1$, with the extremes representing classical and PR box theories, respectively. Putting $\upsilon_{\rm loc}^\ast=2$ and $1+ \frac{1}{\sqrt{2}}$ in Eq. (\ref{eq:bellnonloc}), with $\varsigma=1$, we obtain the Bell-CHSH inequality for PR boxworld and QM, respectively, whilst putting $\upsilon_{\rm loc}^\ast=2$ with $\varsigma=\frac{1}{2}$, we obtain that for classical theory. 

\paragraph{Reality and relativity.}
In a quantum violation of inequality Eq.  (\ref{eq:eta}),  Alice's measurement and her remote-preparation of Bob's state are operationally  well defined and evidently  spacelike-separated. This can be experimentally tested, but is already implicit in loophole-free tests of quantum steering such as \cite{wittmann2012loophole}. Yet, there can't be any operational mechanism causally linking these two events, QM being non-signaling.  On the other hand, there is an  asymmetry and natural causal ordering in the experiment-- namely, that Alice's measurement results in Bob's particle's remote preparation, rather than the other way round. (By contrast, in a Bell test, which is symmetric, the case for  such intrinsic causal ordering is less compelling.)  The only operational element available encompassing both events in spacetime is the probability field corresponding to the quantum wave function, $\psi(x)$. Minimalistically, we attribute to this field itself the capacity to act as a kind of atemporal causal matrix (cf. \cite{gisin2009quantum}). The quantum state is arguably real to possess such causal efficacy. A similar attribution  of reality can be made for states in any GPT in \textbf{Ver}-complete.


In light of operational nonlocality, it is natural to ask why QM, a theory in \textbf{Ver}, isn't found in \textbf{Det}. To clarify this question, we note that in complexity theory, there are sound mathematical grounds to believe that intractable problems exist,  and therefore that $\textbf{P} \ne \textbf{NP}$ \cite{aaronson2013quantum}. In other words, there seems to be a computational barrier separating \textbf{P} and \textbf{NP}. If QM lay outside \textbf{Ver} (like classical theory), then the above query wouldn't be well motivated. To rephrase the question: is there a natural barrier that precludes theories in \textbf{Ver} from living in \textbf{Det}? 

An obvious response would be to invoke the special relativistic prohibition on superluminal signaling. But, the fact is that quantum and classical relativistic no-signaling are fundamentally different. The former corresponds to $ \overline{\textbf{Det}}-\overline{\textbf{Ver}}$ and is a consequence of the tensor product structure of multipartite quantum systems, whereas the latter corresponds to $ \overline{\textbf{Ver}}$ and is an axiom of spacetime geometry. Indeed, even non-relativistic QM is non-signaling. In other words, one would like to look for an \textit{information theoretic} basis for quantum no-signaling in the GPT framework. This would potentially provide a more natural justification for the $\textbf{Ver}/\textbf{Det}$ barrier.

An argument, that we elaborate elsewhere, is the following. It is known that no-signaling can be used to derive various no-go theorems for quantum cloning \cite{gisin1998quantum}, state discrimination \cite{croke2008nosignaling}, etc. Here, we wish to invert this argument  in the context of GPTs. Let us consider a simple 2-input-2-output scenario that illustrates the basic intuition. Correlation $P(a,b|x,y)$ is nonlocal iff the probability $p_{\rm success}$ to satisfy the CHSH condition $a + b = xy \mod 2$ exceeds $\frac{3}{4}$, with $x, y, a, b \in \{0,1\}$. Assume that Alice and Bob share a PR box state $|\textit{PR})_{AB}$. If superluminal signaling were possible, then Bob could  access Alice's input $x$ and prepare a second gbit $B^\prime$ in the state $|b^\prime = a+xy^\prime)_{B^\prime}$, with $a$ obtained via pre-shared randomness. Thus, particle $B^\prime$ would be a clone of $B$, which would contravene the local no-cloning rule for gbits. 

More generally, let spaces $\Omega_A$ and $\Omega_B$ of GPT $\theta$ be non-simplicial (and thus have nonclassical features such as no-cloning \cite{hardy1999no, plavala2016all, aravinda2017origin, barnum2006cloning}), and joint space $\Omega_{AB}$ contain elements outside the minimal tensor product. The above argument suggests that if theory $\theta$ admits signaling, then it would lack  a natural trace operation on $\Omega_{AB}$ to recover the subsystem nonclassicality. 

\paragraph{Conclusions.} An operational concept of locality stronger than signal locality is formulated in terms of a class of operational decision problems, and EPR steering  is proposed to realize it. This possibility is particularly tied to its asymmetric character, which  Bell nonlocality lacks.  Quantum mechanics, and some other operational theories, are shown to be nonlocal according to this operational criterion.

As regards practical demonstration, evidently experimental tests of uncertainty reduction via steering can be adapted to a test of operational nonlocality. Other phenomena, such as those mentioned in the Introduction as evocative of remote disturbance, could be used to propose other versions of operational nonlocality. Also, the argument for the \textit{insecurity} of quantum bit commitment \cite{mayers1997unconditionally, lo1998why, CDP+13} can presumably be reconstructed as a criterion for operational locality, indeed one that, unlike the present one, identifies PR box world as operationally nonlocal.

Steering corresponds to strong correlations between incompatible observables of two particles, leading to the violation of the single-system uncertainty relation. Einstein \textit{et al.} \cite{einstein1935can} believed that this violation indicated QM to be incomplete, since the alternative would be a nonlocal influence that they (wrongly) deemed contradicted by relativity. Here, by grounding this nonlocality in an operational setting, we find this alternative inevitable. In a sense, perhaps we have simply only drawn attention to ``an elephant in the room''.

\acknowledgments

The author thanks S. Aravinda, U. Shrikant and N. Vinod  for discussions, and  AMEF, Bengaluru and DST-SERB,  Govt.  of  India (project EMR/2016/004019) for  financial support.  

\bibliography{qvanta}

\appendix

\section{I. Generalized probability theories (GPTs) \label{sec:gpt}}

In the framework of GPTs,  set $\Omega$ of states $\omega$ of a theory is a  convex  subset of a real vector space $V$.  A valid measurement or observable is any affine functional that maps any state $\omega \in \Omega$ to a probability distribution over measurement outcomes (usually taken to be finite in number). The unnormalized states of the theory form a  convex  positive  cone living  in $V$.

For a bipartite system satisfying the  assumptions  of   no-signaling  and  tomographic  locality \cite{hardy2011reformulating}, the set $\Omega_{AB}$ of all bipartite states lies between (according to a natural ordering) the minimal tensor product $\Omega_A \otimes_{\rm
	min}  \Omega_B$, which denotes the  collection  of all  possible separable  states
(having the form $\sum_j p_j \omega_A^j \otimes \omega_B^j$, where $p_j$ is a probability distribution) and the maximal tensor product
$\Omega_A \otimes_{\rm max} \Omega_B$.

\section{II. No-signaling \label{sec:nosig}}

Suppose Alice's
and  Bob's  correlations  are   described  by  the  conditional  probability distribution
$P(a,b|x,y)$, where  $(a,x)$ (resp., $(b,y)$) are  the (output, input)
pair of  Alice (resp., Bob), and  $a, b, x, y$  are drawn respectively
from  the sets  $\mathcal{A}, \mathcal{B},  \mathcal{X}, \mathcal{Y}$.

Signal locality requires that
\begin{subequations}
\begin{align}
\forall_{y \ne y^\prime} P(a|x,y) &= P(a|x, y^\prime) \equiv 
P(a|x), \label{eq:nosiga} \\
\forall_{x \ne x^\prime} P(b|x,y) &= P(b|x^\prime, y) \equiv P(b|y),
\label{eq:nosigb}
\end{align}
\label{eq:nosig}
\end{subequations}
for any given state shared between  Alice and Bob. 

The dimension of a non-signaling theory (i.e., the number of independent real numbers required to describe an arbitrary mixed state in the theory) is:
\begin{equation}
\mathcal{D}_{\rm   nosig}   =    |\mathcal{X}|   \cdot   |\mathcal{Y}|
(|\mathcal{A}|-1) (|\mathcal{B}|-1)  + |\mathcal{X}| (|\mathcal{A}|-1)
+ |\mathcal{Y}| (|\mathcal{B}|-1)
\label{eq:snosig}
\end{equation}
which can be derived as follows. The last two terms in the RHS of Eq.
(\ref{eq:snosig})  come from  the  marginal probability  distributions $P(a|x)$ and $P(b|y)$ in Eq. (\ref{eq:nosig}), whilst the first  term in the RHS comes from the fact that  for  each  of  the  $|\mathcal{X}|\cdot|\mathcal{Y}|$  pairs  of two-party   inputs,  there   are  $(|\mathcal{A}|-1)(|\mathcal{B}|-1)$
independent output pairs.

In Eq. (\ref{eq:nosig}), consider  the top equation (\ref{eq:nosiga}). For each input $x=X$, given output $a$, there are $|\mathcal{Y}|-1$ independent constraints  by varying  $y$, and  then $|\mathcal{A}|-1$  independent values   to  set   $a$ (minus 1 for normalization).   This   gives  a   total  of   $|\mathcal{X}| (|\mathcal{Y}|-1)    (|\mathcal{A}|-1)$   no-signaling    constraints. Repeating the similar exercise for Eq. (\ref{eq:nosigb}), in all there are $$C_{\rm nosig} = |\mathcal{X}| (|\mathcal{Y}|-1) (|\mathcal{A}|-1) +  |\mathcal{Y}| (|\mathcal{X}|-1)(|\mathcal{B}|-1)$$ constraints (see main text).

For  theories  in  {\bf  Det},  relaxing  the  no-signaling  constraints
(\ref{eq:nosig}), one obtains  $$D_{\rm  prob}   =  D_{\rm   nosig}  +   C_{\rm  nosig}   =
|\mathcal{X}|\cdot|\mathcal{Y}|  (|\mathcal{A}|\cdot|\mathcal{B}|-1),$$
the full dimensionality of a probability polytope.

\section{III. Status of operational locality in various theories
\label{sec:various}}

We survey  various nonclassical operational theories,  besides quantum
mechanics,  concerning  their  operationally nonlocal  behavior. 

\paragraph{Classical   theory:} 
is operationally local since $\upsilon_{\rm  loc}^\ast = 1$ for any pair
of  observables,  implying  that  Eq.   (\ref{eq:eta})  can  never  be
violated.

\paragraph{Quantum mechanics:} 
is operationally nonlocal, as already shown in the main text.

\paragraph{Spekkens' toy theory} \cite{spekkens2007evidence} is operationally nonlocal. The state space of a single system is characterized by four ontic states, labelled 1, 2, 3 and 4. The only single-system pure states are the ``eigenstates''  of three dichotomic and mutually unbiased measurements, $\sigma_{X;{\rm Sp}} \equiv \{1 \vee  2, 3 \vee 4\}, \sigma_{Y;{\rm Sp}} \equiv \{1 \vee 3, 2 \vee  4\}$ and $\sigma_{Z;{\rm Sp}} \equiv \{1 \vee 4, 2  \vee 3\}$.

Without loss of generality, suppose $b \in \{\sigma_{X:{\rm Sp}}, \sigma_{Z;{\rm Sp}}\}$. If  the measured state is an  eigenstate, its outcome can be deterministically predicted, whereas if it is not, then it can be predicted only half the  time.  Thus, $\upsilon_{\rm  loc}^\ast = \frac{3}{2}$.  On  the  other  hand, the lhs of Eq. (\ref{eq:eta}) evaluates to 2 because the theory admits perfect steering (see below), which is related to the incompatibility of any pair of its measurements.

Bipartite entangled states in the toy theory have the form
\begin{equation}
(a \wedge e) \vee (b \wedge f) \vee (c \wedge g) \vee (d \wedge h),
\end{equation} 
where $a, b, c, d, e, g, h \in  \{1, 2, 3, 4\}$ such that $a \ne b \ne
c\ne d$ are ontic states of the first particle, and $e \ne f \ne g \ne
h$ are  ontic states of the  second particle.  To see  that the theory
allows steering, consider the  entangled state with $a = e  = 1$, $b =
f=2$, $c=g=3$ and $d=h=4$. Measuring  in $\sigma_{X;{\rm Sp}}$ (resp., $\sigma_{Z;{\rm Sp}}$) basis, Alice collapses Bob's  state into the corresponding eigenstate of $\sigma_{X;{\rm Sp}}$  (resp., $\sigma_{Z;{\rm Sp}}$).   There is  no  transgression of  no-signaling since  the uniform mixture of $1 \vee 2$ and $3 \vee 4$ equals that of $1 \vee
3$ and $2 \vee  4$, which is the fully mixed state  in the theory.  This
perfect steering behavior implies that the lhs of Eq. (\ref{eq:eta}) evaluates to 2.

However,  the  theory doesn't  admit  states  that are  Bell-nonlocal, because any pair of measurements, although incompatible in the operational theory, are ``meta-compatible'', i.e., the outcome statistics admits a joint distribution (JD); see Appendix IV. Compatibility in the operational theory is thwarted because the required master observable is not part of the theory's set of allowed measurements. This peculiarity has to do with the non-convexity of the theory (arbitrary convex combinations of pure states are not part of the state space $\Omega$).

\paragraph{Generalized local  theory:} 
Single system  states   are   \textit{gdits}, characterized by $d$ fiducial measurements with $k$ outcomes each \cite{barrett2007information}. The state space is the convex hull of the $|\Omega| = k^d$ pure states of single systems, which correspond to deterministic outcomes for each fiducial measurement. The dimension $\textrm{dim}(\Omega)$ of the system is $d(k-1)$, the number of parameters required to describe $d$ probability distributions. Nonclassicality arises from the fact that $\Omega$ is not a simplex, noting that $|\Omega|-\textrm{dim}(\Omega) >1$. The joint space of a multi-partite system is the minimal tensor product $\otimes_{\rm min}$, namely the set of convex  combinations of the direct product of two single-system  states. Therefore, the states are unsteerable and consequently operationally local. 

\paragraph{Boxworld:}    
The single-system states are two-dimensional gdits, namely gbits. By convention, $x, y, a, b \in \{0,1\}$. Pure bipartite entangled states are PR boxes $P(x,y|a,b)$, characterized  by $x  \oplus y  = a\cdot b$, which ensures the maximal violation of the CHSH inequality, while $P(0|a) = P(0|b) = \frac{1}{2}$, ensuring no-signaling \cite{barrett2005popescu}.  Clearly, the lhs of Eq. (\ref{eq:eta}) evaluates to 2, the maximal possible value. Yet, the theory is not operationally nonlocal, because gbits are maximally certain, i.e., $\upsilon_{\rm loc}^\ast=2$. 

In particular, gbits (see Figure \ref{fig:gdit}) take simultaneous deterministic values for $\sigma_{X;g}$ (corresponding to $x,y=0$) and $\sigma_{Z;g}$ (corresponding to $x,y=1$), the gbit analogues of Pauli $\sigma_X$ and $\sigma_Z$. The state space $\Omega_g$ is the convex hull of four pure points, denoted $(0,0), (0,1), (1,0)$ and $(1,1)$, where the first (resp., second) coordinate represents the probability to get outcome 0 if $\sigma_{X;g}$ (resp., $\sigma_{Z;g}$) is measured. Thus, $\upsilon_{\rm loc}^\ast=2$. 

The state space is non-simplicial, and hence features measurement disturbance \cite{aravinda2017origin}, so that only one of the two measurement values can be read out, while the other is maximally disturbed. Therefore, the gbit  pure states in the theory can't be prepared by direct measurement, but instead by measuring one of two particles in a PR box state, which prepares the partner particle in a gbit pure state.

\section{IV. Complementarity in Spekkens' toy theory \label{sec:spekkens}}

Consider two observables in Spekkens' toy theory, say  $\sigma_{X;{\rm Sp}}$ and $\sigma_{Z;{\rm Sp}}$, the analogues of Pauli $\sigma_X$ and $\sigma_Z$. We construct the joint measurement $M$ as a ``master effect'' such that $\sum_j M[j,k]$ reproduces $\sigma_{Z;{\rm Sp}}[k]$, the effect  that corresponds to outcome $k$  on measuring $\sigma_{Z;{\rm Sp}}$, and similarly $\sum_k M[j,k] = \sigma_{X;{\rm Sp}}[j]$. Let $x^{\pm1}$ be the vector representing the pure (maximum information) states of measurement $\sigma_{X;{\rm Sp}}$, and $z^{\pm1}$ those of measurement  $\sigma_{Z;{\rm Sp}}$, and so on. Now,
\begin{align}
M[++].x^+ &+ M[+-].x^+ = 1 \nonumber\\
M[-+].x^+ &+ M[--].x^+ = 0 \nonumber\\
M[++].x^+ &+ M[-+].x^+ = 0.5 \nonumber\\
M[+-].x^+ &+ M[--].x^+ = 0.5
\end{align}
from which it follows that 
\begin{eqnarray}
M[++]\cdot x^+ &=& M[+-]\cdot x^+ = 0.5 \nonumber\\
M[-+]\cdot x^+ &=& M[--]\cdot x^+ = 0.
\label{eq:cp-}
\end{eqnarray}
Proceeding thus, one finds
\begin{align}
M[++]\cdot x^- &= M[+-]\cdot x^- = 0 \nonumber \\
M[-+]\cdot x^- &= M[--]\cdot x^- = 0.5 \nonumber \\
M[+-]\cdot z^+ &= M[--]\cdot z^+ = 0 \nonumber \\
M[++]\cdot z^+ &= M[-+]\cdot z^+ = 0.5 \nonumber \\
M[++]\cdot z^- &= M[-+]\cdot z^- = 0 \nonumber \\
M[+-]\cdot z^- &= M[--]\cdot z^- = 0.5.
\label{eq:cp}
\end{align}
The nonclassicality turns up in the fact that the extreme points
are not linearly independent. In particular, 
\begin{equation}
x^+ + x^- = z^+ + z^-.
\label{eq:li}
\end{equation} 
It may be verified that each  of the components $M[j,k]$ determined in Eqs. (\ref{eq:cp-}) and (\ref{eq:cp}) are consistent with  Eq. (\ref{eq:li}).  Thus, in this toy theory, $\sigma_{X;{\rm Sp}}$ and $\sigma_{Z;{\rm Sp}}$, and by a similar argument any pair of observables admits a JD. Therefore, although the pairs are incompatible in the operational theory, they are ``meta-compatible'', i.e., compatible in an underlying HV model. Meta-compatiblity entails that a joint probability distribution  exists for  all  measurement outcomes in the two-party Bell-CHSH scenario. Therefore, by Fine's theorem \cite{fine1982hidden}, the correlations must be local.

\section{V. Uncertainty and steering strength bound on Bell nonlocality}

The interplay of uncertainty and steering brought out by our result casts light on the bound on Bell nonlocality from (fine-grained) uncertainty and steering in a non-signaling theory \cite{oppenheim2010uncertainty}. (In a signaling theory, the signal directly demonstrates nonlocality, even without uncertainty and steering.)  Further comments concerning Bell nonlocality are in order here.

Like in operational nonlocality, incompatibility is necessary in Bell nonlocality (although the specific settings for maximal violation can be different). In the two-input-two-output case with outputs $\pm1$, suppose   $y_0$ and  $y_1$ are  compatible.   Measuring $\tilde{\omega}_B^{a_0|x_0}$  (resp., $\tilde{\omega}_B^{a_1|x_1}$) in bases $y_0$ and $y_1$ yields JD $P(b_0, b_1,  a_0~|~y_0,  y_1,  x_0)$  (resp., $P(b_0,  b_1,  a_1~|~y_0,  y_1, x_1)$).  One  can then construct  JD $P(a_0, a_1, b_0, b_1~|~x_0, x_1, y_0, y_1)$ given by 
\begin{equation}
\frac{P(b_0,  b_1,  a_0~|~y_0, y_1, x_0)P(b_0,  b_1,
	a_1~|~y_0, y_1, x_1)}{P(b_0, b_1~|~y_0, y_1)},
\label{eq:fine}
\end{equation}
which reproduces the observed JD's by tracing over $a_0$ or $a_1$.
By  Fine's theorem  \cite{fine1982hidden},  such  a correlation  can't violate  a Bell  inequality, and thus  must satisfy the CHSH locality condition  
\begin{equation}
|\langle  a_0b_0\rangle   +   \langle a_0b_1\rangle + \langle a_1b_0\rangle  - \langle a_1b_1\rangle| \le 2,
\label{eq:CHSHc}
\end{equation}
which we obtained from only arguments related to compatibility, rather than local-realism  (cf. \cite{son2005joint}).
Eq. (\ref{eq:CHSHc})  is maximally violated for $a \in \{\frac{\sigma_X \pm \sigma_Z}{\sqrt{2}}\}$ and $b \in \{\sigma_X, \sigma_Z\}$, with $\omega_{AB}$ being the singlet state. On the other hand, we saw that maximal qubit violation of Eq. (\ref{eq:eta}) requires setting $a=b$.

In the context of GPTs, incompatibility  can be quantified in terms of an  ``unsharpness'' parameter,  $\kappa$: given dichotomic   observable   $O$,  its   unsharp  version   is defined to be $O^{(\kappa)} = \kappa O  + (1-\kappa)\frac{I}{2}$, where $(0 <  \kappa \le  1)$ and $I$ is the uniform distribution over two inputs \cite{busch2013comparing}. It follows  that  the expectation  value  $\langle  a_j b_k^{(\kappa)}\rangle  = \kappa  \langle a_j  b_k\rangle$.  The degree of compatibility is the maximum   $\kappa$    such   that $b_0^{(\kappa)}$ and $b_1^{(\kappa)}$  are jointly measurable. Let $\kappa_{\rm opt}$ denote such a maximum, optimized over all pairs of  measurements. It follows that the CHSH inequality Eq.  (\ref{eq:CHSHc}), with Alice and Bob measuring $a_j$ and $b_k^{(\kappa_{\rm opt})}$, takes the form 
\begin{eqnarray}
|\langle  a_0b_0\rangle  +  \langle a_0b_1\rangle  +
\langle a_1b_0\rangle - \langle  a_1b_1\rangle| \le \frac{2}{\kappa_{\rm opt}},
\label{eq:baniq}
\end{eqnarray}
an incompatibility-based bound on nonlocality \cite{banik0degree}. 

To relate compatibility $\kappa_{\rm   opt}$ to uncertainty, we observe that roughly speaking, among GPTs  with the same  number of  inputs and outputs, a more uncertain theory tends to feature more compatibility (also, cf. \cite{sun2018uncertainty}). For example, in the 2-input-2-output case, PR boxworld features  the least compatibility ($\kappa_{\rm opt}=\frac{1}{2}$) and zero  uncertainty ($\upsilon_{\rm loc}^\ast=1$), whereas in Spekkens theory all pairs of measurements are  meta-compatible (as noted earlier) and  uncertainty is maximal (all measurements are mutually  unbiased).  QM  occupies  an intermediate  position here. 

This pattern may be  understood as follows. Among these GPTs, gdit theory can be considered as providing an ontological model,  with states and measurements in the other theories, having greater uncertainty, being considered as (ontologically) more smeared  versions of those in gdit theory (see Figure \ref{fig:gdit}, Appendix VI). Thus, measurements in a more uncertain theory require lesser  fuzzification  to attain joint measurability. 

For example, consider a family of 2-input-2-output GPTs $\theta_\tau$ such that $\lambda^\tau + \mu^\tau\le 1$ ($\tau \ge 1$) determines the parameter region $\Delta$ where observables $\sigma_{X;\tau}^{(\lambda)}$ and $\sigma_{Z;\tau}^{(\mu)}$ (the $\theta_\tau$ analogues of $\sigma_X$ and $\sigma_Z$, with unsharpness parameters $\lambda$ and $\mu$, respectively) are jointly measurable. Larger  $\tau$ represents a greater area of $\Delta$ and hence \cite{busch2013comparing} greater compatibility. We set $\kappa_{\rm opt}$ to be the maximum $\lambda$ such that $\lambda=\mu$, giving $\kappa_{\rm opt}=2^{-(1/\tau)}$ for theory $\theta_\tau$, and the Tsirelson bound in Eq. (\ref{eq:baniq}) becomes $2\cdot2^{\frac{1}{\tau}}$. Here, $\tau=1$ and 2 yield the bound for  PR boxworld and QM, respectively  \cite{stevens2014steering}.  For theories $\theta_\tau$ and their more classical counterparts, we find (Appendix VI):
\begin{equation}
\kappa_{\rm opt} = \frac{\beta}{2(\upsilon_{\rm loc}^\ast-1)},
\label{eq:uncertain}
\end{equation}
with the allowed range being $\frac{1}{2} \le \kappa_{\rm opt} \le 1$. Eq. (\ref{eq:uncertain})
shows that more compatible theories are more uncertain, for fixed $\beta$  (see above). Here, family parameter $\beta~ (\in [1,2])$ interpolates between the family $\theta_\tau$ (characterized by a non-simplicial state space $\Omega$; $\beta=1$)  and classical theory (simplicial $\Omega$; $\beta=2$), and the range of $\upsilon_{\rm loc}^\ast$ is determined by the above allowed range of $\kappa_{\rm opt}$. Importantly, unlike incompatibility, uncertainty must be combined with parameter $\beta$ to capture nonclassicality: for example, $\upsilon_{\rm loc}^\ast=1$ can refer to both classical theory ($\kappa_{\rm opt}=1$) and gbit theory ($\kappa_{\rm opt}=\frac{1}{2}$). 

Using Eq.  (\ref{eq:uncertain}) in Eq. (\ref{eq:baniq}), we obtain the inequality (\ref{eq:bellnonloc}), where we have set $\varsigma \equiv \frac{1}{\beta}$. By virtue of our earlier result linking incompatibility and steering, $\varsigma$ represents the degree of steering, i.e., greater $\varsigma$ corresponds to lesser classicality and more steering. 

\section{VI. Incompatibility and uncertainty for PR boxworld, QM and a family of theories}
\begin{figure}
	\begin{center}
		\begin{tikzpicture}[scale=2]
		\draw (0,0) rectangle (1,1);
		\draw [dashed] (0,0.5) -- (1,0.5);
		\draw [dashed] (0.5,0) -- (0.5,1);
		\draw (0,1) -- (1,0);
		\draw (0,0) -- (1,1);
		\draw (1,1) circle node [black,above] {(1,1)};
		\draw (0,0) circle node [black,below] {(0,0)};
		\draw (0,1) circle node [black,above] {(0,1)};
		\draw (1,0) circle node [black,below] {(1,0)};
		\draw (0.5,0.5) circle [radius=0.5];
		\end{tikzpicture}
		\caption{State space of gdit theory,  fragment of QM (with $b_0=\sigma_X$ and
			$b_1=\sigma_Z$)  and of  Spekkens' toy  theory: the  vertices of  the solid
			square represent the four extreme states of the gdit theory, denoted
			by  $(P(+1|b_0),  P(+1|b_1)$.   Qubits  are  represented  by  the
			inscribed  circle (see  Figure 2  in \cite{barrett2007information}).
			The dashed  lines represent Spekkens' theory (Appendices III and IV). The
			point where the diagonal lines intersect the circle, which represent
			the quantum states with  maximum uncertainty,  are eigenstates  of $(\sigma_X  \pm
			\sigma_Z)/\sqrt{2}$. All these theories are  nonclassical in that the state
			space  is non-simplicial  \cite{plavala2016all,aravinda2017origin},
			i.e., their extreme points aren't linearly independent. Note that gdit theory can be considered as providing an ontological model for the other theories.}
		\label{fig:gdit}
	\end{center}
\end{figure}
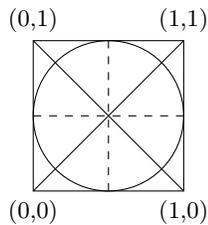

Figure \ref{fig:gdit} depicts the idea that gbit theory can be considered as an ontological model for a family of other 2-input-2-output theories with non-vanishing uncertainty. Essentially, we regard observables and states in these theories as the smeared or fuzzified versions (at the ontological level) of their gbit counterparts. The family of GPTs $\theta_\tau$ is a simple quantitative illustration of this idea.

Let $\sigma_{X;g}$ and $\sigma_{Z;g}$ be the analogues of Pauli $\sigma_X$ and $\sigma_Z$ in gbit theory. The corresponding observables in theory $\theta_\tau$ are denoted $\sigma_{X;\tau}$ and $\sigma_{Z;\tau}$, and their unsharp versions by $\sigma_{X;\tau}^{(\lambda)}$ and $\sigma_{Z;\tau}^{(\mu)}$. Theory $\theta_\tau$ is characterized by region $\Delta$ of joint measurability (in the ($\lambda,\mu$)-parameter space) of $\sigma_{X;\tau}^{(\lambda)}$ and $\sigma_{Z;\tau}^{(\mu)}$ given by 
\begin{equation}
\lambda^\tau + \mu^\tau \le 1,
\label{eq:taufamily}
\end{equation}
where $0 < \lambda, \mu \le 1$ and $\sigma_{X;\tau}$. The larger is $\tau$, the larger is the area of $\Delta$, and correspondingly more compatibility in the theory, according to the criterion proposed in \cite{busch2013comparing}. A measure of compatibility, which can be identified with $\kappa_{\rm opt}$, is obtained by setting $\lambda=\mu$ such that Eq. (\ref{eq:taufamily}) is saturated. Accordingly, we find:
\begin{equation}
\kappa_{\rm opt} = 2^{-1/\tau},
\label{eq:mutau}
\end{equation}
which, for gdit theory, assumes the minimum value
\begin{equation}
\kappa_{\rm opt}^{\rm gdit} = \frac{1}{2},
\label{eq:mug}
\end{equation}
setting $\tau=1$. 

Suppose $\sigma_{X;\tau} = \alpha \sigma_{X;g} + (1-\alpha)(I/2)$, by the (ontological) smearing of the corresponding gbit observable, where $\alpha$ is the ontological unsharpness parameter. Smearing $\sigma_{X;\tau}$ at the operational level, we obtain $\mu \sigma_{X;\tau} + (1-\mu)\sigma_{X;\tau} = \mu\alpha \sigma_{X;g} + (1-\mu\alpha)(I/2)$. Therefore the parameter $\mu\alpha$, which is like the effective smearing at the ontological gbit level, must satisfy
\begin{equation}
\mu\alpha=\frac{1}{2},
\end{equation}
in view of Eq. (\ref{eq:mug}). Letting $\mu$ $\equiv \kappa_{\rm opt}$ in Eq. (\ref{eq:mutau}), we find:
\begin{equation}
\alpha = 2^{-1 +(1/\tau)},
\label{eq:mualfa}
\end{equation}
which represents the required ontological smearing with respect to gdit theory, to reproduce a $\theta_\tau$ observable.

We can now estimate uncertainty parameter $\upsilon_{\rm loc}^\ast$ in $\theta_\tau$ as follows. Suppose that the action of $\sigma_{X;\tau}$ with respect to some state in this GPT is captured by the above smearing, but, conservatively speaking, $\sigma_{Z;\tau}$ is not, so that the output of $\sigma_{Z;\tau}$ on this state is deterministic as in gbit theory. Thus:
\begin{eqnarray}
\upsilon_{\rm loc}^\ast &=& 1+\alpha \nonumber\\
&=& 1+2^{-1 +(1/\tau)} \nonumber\\
&=& 1 + \frac{1}{2\kappa_{\rm opt}},
\label{eq:upsonto}
\end{eqnarray}
as the relation between uncertainty and incompatibility in the $\theta_\tau$ GPT family. Cases $\tau=1$ and $\tau=2$ give the known bounds for gdit theory and QM \cite{stevens2014steering}.

Classical theory, like PR boxworld, lacks uncertainty, but unlike in the latter, the state space is a simplex. Thus, it does not belong to the above $\theta_\tau$ family and cannot be derived from by an ontological fuzzification of gbit theory. This is reflected, for example, in the fact that setting $\kappa_{\rm opt} := 1$ in Eq. (\ref{eq:upsonto}) doesn't lead to full certainty. Classical theory can be incorporated into this scheme, by letting space $\Omega^{\rm classical}$ be obtained from gdit space $\Omega^{\rm gbit}$ by ``simplexifying''  the latter. This is a bijective map that ``inflates" the outer square in Figure \ref{fig:gdit} to a tetrahedron (cf. \cite{aravinda2017origin}). For our present purpose, we can capture this process by a parameter $\beta$ ($1 \le \beta \le 2$) that modifies Eq. (\ref{eq:upsonto}) to
\begin{equation}
\upsilon_{\rm loc}^\ast = 1 + \frac{\beta}{2\kappa_{\rm opt}},
\label{eq:gditclassical}
\end{equation}
from which Eq. (\ref{eq:uncertain}) follows. Here, $\beta=1$ corresponds to the $\theta_\tau$ family of nonclassical theories derived from gdit theory, whilst $\beta=2$ corresponds to classical theory. Thus, $\beta$ is like a family parameter interpolating between classical theory and the $\theta_\tau$ family. In Eq. (\ref{eq:gditclassical}), both gdit theory ($\kappa_{\rm opt}=\frac{1}{2}, \beta=1$, i.e., minimal compatibility and non-simplicial $\Omega$) and classical theory ($\kappa_{\rm opt}=\frac{1}{2}, \beta=1$, or maximum compatibility and simplicial $\Omega$) are seen to correspond to vanishing uncertainty.

\end{document}